# Reference pulse attack on continuous variable quantum key distribution with local local oscillator under trusted phase noise

SHENGJUN REN[1],*, RUPESH KUMAR[2], ADRIAN WONFOR[1], XINKE TANG[1], RICHARD PENTY[1], AND IAN WHITE[1]

[1]*Centre for photonic systems, University of Cambridge CB3 0FA, UK*
[2]*Quantum Communications Hub, University of York,, YO10 5DD, UK*
*\*sr734@cam.ac.uk*

**Abstract:** We show that partially trusting the phase noise associated with estimation uncertainty in a LLO-CVQKD system allows one to exchange higher secure key rates than in the case of untrusted phase noise. However, this opens a security loophole through the manipulation of the reference pulse amplitude. We label this as 'reference pulse attack' which is applicable to all LLO-CVQKD systems if the phase noise is trusted. We show that, at the optimal reference pulse intensity level, Eve achieves unity attack efficiency at 23.8km and 32.0km while using lossless and 0.14dB/km loss channels, respectively, for her attack. However, in order to maintain the performance enhancement from partially trusting the phase noise, countermeasures have been proposed. As a result, the LLO-CVQKD system with partially trusted phase noise owns a superior key rate at 20km by an order 9.5, and extended transmission distance by 45%, than that of the phase noise untrusted system.

## 1. Introduction

The aim of quantum key distribution (QKD) is to promise theoretically validated information security between two authenticated users, Alice and Bob. This is achieved by sharing common random secure keys which are unknown to an eavesdropper-Eve [1-4]. Continuous Variable (CV) based QKD has rapidly developed in recent years [5-7] and its protocols take advantage of the properties of light associated with its wave nature. Here keys are encoded on amplitude and phase jointly referred as quadrature, of the light which can be extracted by shot noise limited coherent detection. The primary advantage of CVQKD is its compatibility with standard off-the-shelf optical communication components which could enable more affordable QKD networks [8]. In addition, while comparing with dedicated single-photon detector based discrete variable (DV) QKD systems, CVQKD systems are efficient and have higher secure key rates over access and metro dense wavelength division multiplexed (DWDM) networks [9-11]. Amongst number of CVQKD protocols, Gaussian-modulated coherent-state (GMCS) CVQKD having been rigorously studied and shown to provide unconditional security against malicious eavesdropping attacks [12,13]. However, imperfections in real CVQKD systems result in potential loopholes that compromise the secure key generation. For example, attacks that exploit the intensity fluctuations of the local oscillator (LO) [14], the wavelength dependency of the homodyne beam splitter [15], and the saturation of the homodyne detector [16] have been identified and respective counter measures have been proposed.

The Local Local Oscillator (LLO) CVQKD system, proposed and demonstrated in [17-19], obviates the need for direct transmission of a LO and thereby nullifies the scope for direct or indirect attacks on the LO. This is achieved by using independent narrow linewidth lasers of the same center wavelength, one at Alice for signal generation and the other at Bob for local oscillator use. However, agreeing a common phase reference between the two free running lasers is an experimental challenge. In order to ease such a limitation, Alice can share low intensity phase reference pulses with Bob, that can either lock the phase of the LO laser with the one at Alice in real time [20] or otherwise can recover the phase and corrects the measurements later during data reconciliation [21]. The latest LLO implementations based simultaneous pilot's signals [22], polarization multiplexed pilots [23] and more detailed noise analysis [24] also comes to the stage.

Due to the relatively low intensity, the quantum uncertainty associated with the reference pulse's phase estimation induces a phase estimation error in Bob's quadrature measurement. The phase drift due to the laser spectral linewidth also contributes to the phase noise. Conventionally, in order to upper bound the information leakage, it is assumed that phase noise is originated from Eve. This may not be necessary since some part of the phase noise sources can be well identified, calibrated and hence trusted i.e., they are not originated from eavesdropping.

A refined noise model to enhance the phase noise tolerance in LLO-CVQKD has recently been studied in [25] where phase estimation error noise can be locally calibrated and treated as trusted noise while the phase drift noise is still regarded as untrusted. In this model, significant increment in achievable secret key rate and transmission distance has been predicted [17, 21]. Moreover, in [23], we find that the phase variation arise from signal and pilot pulses separation inside Bob is also isolated from Eve, hence certain trusted phase noise and predicted improved performance. Nevertheless, security analysis with trusted phase noise has not been comprehensively investigated and we have therefore found that amplitude-related phase estimation error noise can result in a loophole for Eve to mount an attack.

In this paper, we explore LLO-CVQKD system performance under trusted- referred as realistic, and untrusted- referred as paranoid, security model of phase noise and show the enhancement in secure key rate. And then propose an attack that exploits the phase estimation error associated with the amplitude of the phase reference pulses. The corresponding reduction in phase estimation error leave freedom for Eve to obtain a considerably larger amount of secret key information than that estimated by Alice and Bob, under realistic security model. We call this attack the "reference pulse attack". In the proposed countermeasure, we re-visit the phase noise that can be trusted conditioned on the amount of attack that Eve can mount and then bound her accessible information. As a result, performance is degraded but is better than that of paranoid security model.

This paper is organized as follows. In section II, we briefly describe the LLO-CVQKD system and demonstrate assumption of the trusted phase noise model and performance. In section III, we present the reference pulse attack and we revisit the expression for the mutual information under the reference pulse attack. In section IV, we compare the attack performance under the paranoid and realistic security conditions. We then optimize the trusted phase noise model to provide a countermeasure to the attack. Finally, conclusions are drawn in section V.

## 2. The LLO CVQKD System

In this section, we describe the implementation of the LLO-CVQKD system and the current state-of-art noise model. We focus on the phase noise, especially the phase estimation error noise, to reveal the loophole.

### 2.1 LLO CVQKD principle

As shown in Fig. 1, a practical LLO-CVQKD system using the GMCS protocol can be analysed in four parts: Alice's preparation, channel propagation, Bob's detection and data post processing.

First of all, Alice prepares a train of Gaussian modulated coherent signal states $\alpha_A$ with quadrature $X_A$ and $P_A$ with variance $V_A$ and zero mean. Then she transmits these coherent states to Bob through the quantum channel, which induces an arbitrary phase rotation of the states in the phase space. In order to recover the initial phase of the signal, a phase reference pulse, $\alpha_A^R$ of quadrature $X_A^R$ and $P_A^R$ is transmitted along with each signal state. The intensity of the reference pulse is a few orders of magnitude greater than the signal variance. It is of great importance that the reference pulse amplitude is not too large to limit signal-reference pulse interference. The quantum channel is characterized by the channel parameters T and $\xi_e$, where T is the channel transmittance and $\xi_e$ is the excess noise associated with the action from malicious eavesdropper, Eve.

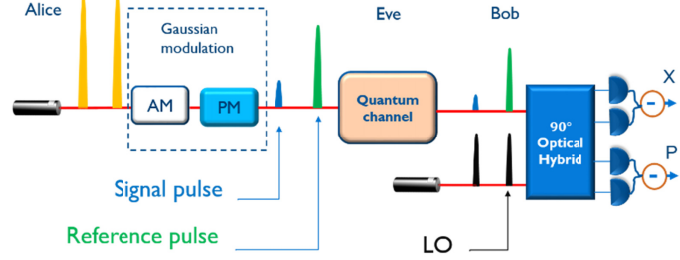

Fig. 1. Practical LLO-CVQKD setup. The signal pulse (blue) and phase reference pulse (green) are transmitted through the optical channel and undergo same attenuation. At Bob, the received signals are interfered with the LO (black) to extract quadrature values.

At Bob, the quadrature of the signal and the reference pulses are measured with intense LO pulses using a shot-noise limited heterodyne detector with an efficiency $\eta$. Unlike the conventional CVQKD implementation, in the LLO-CVQKD system a separate laser at Bob is used to generate the LO. In order to robustly estimate the misalignment of the reference frames of the two free-running lasers, the mean reference pulses' quadrature values at Alice are publically announced. The quadrature measurement outcomes $X_B^R$ and $P_B^R$ of the reference pulse are associated with estimated rotation angle, $\varphi_{esti}^R$, as given below:

$$X_B^R = \sqrt{\frac{T\eta}{2}}(X_A^R \cos\varphi_{esti}^R + P_A^R \sin\varphi_{esti}^R) \qquad (1)$$

$$P_B^R = \sqrt{\frac{T\eta}{2}}(-X_A^R \sin\varphi_{esti}^R + P_A^R \cos\varphi_{esti}^R) \qquad (2)$$

The estimation can be calculated without lose of generality by preparing the initial reference pulse with a zero phase angle (i.e., $P_A^R = 0$)

$$\varphi_{R\_esti} = \tan^{-1}(\frac{P_B^R}{X_B^R}) \qquad (3)$$

Consequently, the estimated phase angle is applied to correct Alice's initial quadrature or Bob's measured quadrature values. This is followed by channel parameter estimation, error correction and privacy amplification.

### 2.2 Trusted phase noise model

However, the phase rotation of the quantum signals cannot be completely corrected through the phase estimation of the reference pulse. As shown in Fig.2, the quantum signal phase rotation $\varphi_S = \theta_{LLO}^S - \theta_{Source}^S$ is the phase difference between the quantum signal at Alice,

$\theta^S_{Source}$, and the local local oscillator, $\theta^S_{LLO}$ at Bob. When reference pulses are deployed to estimate $\varphi_S$. The phase noise in its estimation primarily originates from: (a) the phase drift due to the spectral linewidths $\Delta v_A$, of the laser at Alice, and $\Delta v_B$, of the laser at Bob and (b) the deviation of estimated phase value, $\varphi_{R\_esti}$, of the reference pulse from the exact phase value, $\varphi_R$.

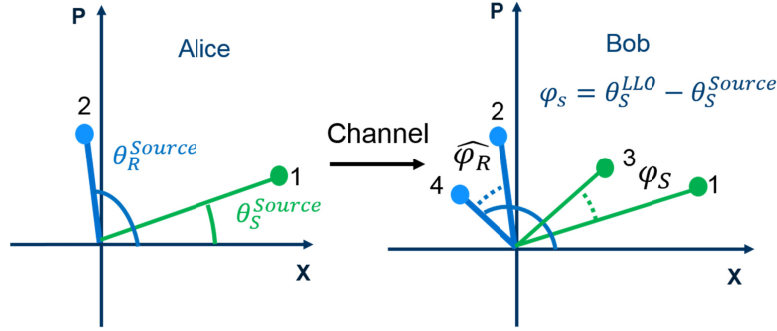

Fig. 2 The general process of phase rotation and estimation. At Bob, the actual relative phase, $\varphi_S$, (green angle) of the quantum signal (1,3) and estimated relative phase $\varphi_R$ (blue angle) from reference pulse (2,4) are added to the initial phase. $\varphi^R_{esti}$ is used to estimate the phase difference $\varphi_S$ between two free-running lasers.

Since the relative phase drift between two free running lasers can be modelled as a Gaussian stochastic process, centred at the central frequency, the variance of this phase drift over the time interval between the signal and reference pulses can be estimated from:

$$V_{drift} = \frac{2\pi(\Delta v_A + \Delta v_B)}{f_{rep}} \quad (4)$$

where $f_{rep}$ is the repetition rate of the system.

Similarly, one can find the uncertainty, $V_{esti}$, associated with the measured and the exact phase of the reference pulse, which is inversely proportional to the amplitude $E_{Ref}$ of the reference pulse at Bob [21]. This is primarily due to the fundamental shot noise and total noise $\varepsilon_{tot}$ which is explicitly defined in next section, and can be written as [18,21]:

$$V_{esti} = \text{var}(\varphi_{R\_esti} - \varphi_R) = \frac{(\chi_{tot}+1)}{E^2_{Ref}} \quad (5)$$

Both these uncertainties, $V_{esti}$ and $V_{drift}$, contributing to phase estimation error noise $\xi_{esti}$ and the drift noise $\xi_{drift}$ and result in a total phase noise $\xi_{phase}$, which can be written as [17,18]:

$$\xi_{phase} \approx V_A * (\frac{\chi_{tot}+1}{E^2_{Ref}} + 2\pi\frac{\Delta v_A + \Delta v_B}{f_{rep}}) \quad (6)$$

In this paper, we adopt the assumption of trusted detector model commonly used in LLO system [17-19, 21, 26] where the noise and efficiency associated with detector are well-protected from Eve. Moreover, we follow the critical trusted phase noise assumption in [25] that the phase estimation error noise $\xi_{esti}$ can be trusted while phase drift noise $\xi_{drift}$ is not. To clarify the assumption, we emphasise that $\xi_{esti}$ is originated from the reference pulse quantum uncertainty at detector - specifically the vacuum noise and experimental noise at detector.

This follows the generally accepted trusted detector noise assumption in practical CVQKD system [24-26]. Meanwhile, $\xi_{esti}$ is determined at the trusted detector and can be locally calibrated at Bob. To achieve pre-calibration, one can use training pulse sessions, as used in classical coherent communications, to estimate total noise and then the calibration of phase estimation noise which can assumed to be constant during the protocol run. Therefore, $\xi_{esti}$ is assumed as trusted detector noise. However, it is hard to verify the randomness of phase drift noise $\xi_{drift}$ associated with two independent lasers [22] so it is assumed as untrusted noise. Based on these two assumptions, the quadrature values measured by Bob under trusted phase noise can be written as:

$$\begin{pmatrix}\hat{X}_B \\ \hat{P}_B\end{pmatrix} = \sqrt{\frac{T\eta}{2}}\left[\begin{pmatrix}\cos\varphi_{R\_esti} & \sin\varphi_{R\_esti} \\ -\sin\varphi_{R\_esti} & \cos\varphi_{R\_esti}\end{pmatrix}\begin{pmatrix}X_A \\ P_A\end{pmatrix} + \begin{pmatrix}X_\xi + X_N \\ P_\xi + P_N\end{pmatrix}\right] + \begin{pmatrix}x_{ele} \\ p_{ele}\end{pmatrix} \quad (7)$$

$X_N$ and $P_N$ are vacuum quadratures with unit shot noise variance ($N_0$). $X_{ele}$ and $P_{ele}$ are electronic noise quadratures with variance $V_{ele}$. In the Eq. (7), the rotation matrix in the first term accounts for the phase estimation error while the $X_\xi$ and $P_\xi$ represent the noise due to Eve's attack.

### 2.3 Trusted phase noise model performance analysis

Secret key formula pertaining to the LLO-CVQKD scheme follows exactly that for a conventional CVQKD scheme. Under a collective attack [3] with reverse reconciliation, the asymptotic secure key rate is:

$$K_{collective} = \beta I_{AB} - \chi_{BE} \quad (8)$$

Where $\beta$ is the reconciliation efficiency and $I_{AB}$ is the mutual information between Alice and Bob while $\chi_{BE}$ is the upper bound of Eve's information related to the Holevo bound [27]. Before we detail the Eq.8 further, we describe the noise nomenclatures used in its expansion. This is important to comprehend the basic difference in the key rate estimation under trusted and untrusted noise model. The trusted phase noise model does not change any derivation procedure rather attribute the trusted noise differently in the parameters involved. This as follows.

As mentioned in previous section, the phase estimation error noise $\xi_{esti}$ in trusted phase noise model is treated as part of the trusted detector noise $\varepsilon_{het}$ while the rest of phase noise – phase drift noise $\xi_{drift}$ and other noises such as AM dynamic range related noise $\xi_{AM}$ [21] and ADC quantization noise $\xi_{ADC}$ [23] are untrusted. The trusted phase noise model does not change the estimation procedure of the eigenvalues. Based on the heterodyne detection for reference and signals, the noise induced by detector under such noise model is expressed as:

$$\chi_{het}^T = \frac{[1 + (1-\eta) + 2V_{ele}]}{\eta} + T\xi_{esti} \quad (9)$$

The total line noise induced in channel containing all untrusted noise referred to channel input is

$$\chi_{line}^T = \frac{1}{T} - 1 + \xi_e + \xi_{drift} + \xi_{AM} + \xi_{ADC} \quad (10)$$

In comparison, overall phase noise is considered to be controlled by Eve in conventional paranoid security model and $\xi_{esti}$ is treated as part of excess noise. Therefore, the channel $\chi^u_{line}$ and the heterodyne detection noise $\varepsilon^u_{het}$, u stands for untrusted, can be obtained as:

$$\chi^u_{line} = \chi^T_{line} + \xi_{esti} \tag{11}$$

$$\chi^u_{het} = \chi^T_{het} - T\xi_{esti} \tag{12}$$

The overall noise referred to the channel input, in trusted noise model, is shown:

$$\chi_{tot} = \chi_{line} + \frac{\chi_{het}}{T} \tag{13}$$

This is also true for untrusted noise case. From the Shannon equation [28], $I_{AB}$ with heterodyne detection can be derived through:

$$I_{AB} = \log_2 \frac{V_B}{V_{B|A}} = \log_2 \frac{V + \chi_{tot}}{1 + \chi_{tot}} \tag{14}$$

Furthermore, the accessed information by Eve $\chi_{BE}$ is evaluated using the Von Neumann entropy S(.) [29]:

$$\chi_{BE} = S(E) - S(E|B) \tag{15}$$

For the GMCS protocol, the Von Neumann entropy is analysed using the symplectic eigenvalues [30-31] of the covariance matrix. The derivation $\chi_{BE}$ in LLO-CVQKD system [17] follows exactly same as that of conventional CV-QKD, with added phase noise from reference pulse as well as laser phase drift. In our estimation, since we have considered the phase estimation error noise is trusted and part of detection noise, it is adequate to follow the same procedure. The mutual information between Eve and Bob is given by:

$$\chi_{BE} = \sum_{i=1}^{2} G\left(\frac{\lambda_i - 1}{2}\right) - \sum_{i=3}^{5} G\left(\frac{\lambda_i - 1}{2}\right) \tag{16}$$

where $G(x) = (x+1)\log_2(x+1) - x\log_2 x$ and the eigenvalues $\lambda_{1,2} = \sqrt{\frac{1}{2}\left(A \pm \sqrt{A^2 - 4B}\right)}$ with

$$A = V^2(1 - 2T) + 2T + T^2(V + \chi_{line})^2 \tag{17}$$

$$B = [T(V\chi_{line} + 1)]^2 \tag{18}$$

Considering that heterodyne detection, $\lambda_{3,4,5}$, can be written by $\lambda_{3,4} = \sqrt{\frac{1}{2}\left(C \pm \sqrt{C^2 - 4D}\right)}$ and $\lambda_5 = 1$ with

$$C = \frac{1}{(T(V+\chi_{tot}))^2}\left[A\chi^2_{het} + B + 1 + 2\chi_{het}\left(V\sqrt{B} + T(V + \chi_{line})\right) + 2T(V^2 - 1)\right] \tag{19}$$

$$D = \left(\frac{V + \sqrt{B}\chi_{het}}{T(V + \chi_{tot})}\right)^2 \tag{20}$$

The simulation for performance comparison is conducted with parameters selected to match with the parameters used in recent LLO-CVQKD experiments [17,19], which are: $V_A = 4$, $\beta = 0.95$, $\eta = 0.5$, attenuation coefficient $\alpha_{std} = 0.2$ dB/km, $V_{elec} = 0.01$, $\xi_e = 0.01$, $f_{rep} = 100$ MHz, $\Delta v = 1.9 kHz$, AM extinction ratio is 40dB and quantization number of the ADC is 11 bits. Among this parameter, most importantly, the intensity of reference pulse needs to be chosen carefully to achieve the best system performance. Too weak reference pulse intensity will introduce large phase noise due to shot noise. On the other hand, even though large reference pulse intensity leads to low phase estimation error noise. However, other noises such as modulation noise due to finite amplitude modulation range and polarization multiplexing device's dynamic range [25], ADC noise [23], etc. will correspondingly increase with reference pulse intensity. The best performance under untrusted phase noise model is around the intensity that expressed in signal variance is $\frac{E_R^2}{V_A} = 100 \sim 150$ in LLO-CVQKD system. We set $\frac{E_R^2}{V_A} = 100$ is as the optimal value in our comparison; also used in ref [25].

As the results shown in Fig. 3, since the trusted phase noise is not regarded as part of the excess noise, it has significantly improved key rate and distance performance to 99.8km. The paranoid security model usually overestimates excess noise, hence Eve's information, resulting in extremely limited system performance to 18.7km. This indicates that relaxations on existing stringent security assumption will lead to better QKD system performance. However, this security assumptions may open new vulnerabilities and potential side channels. Here is no difference; we have identified an attack in trusted LLO-CVQKD system. We explain this in the following section.

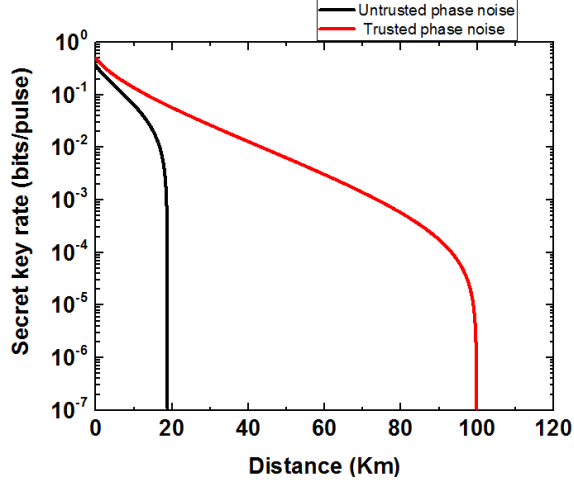

Fig. 3. Secret key rate comparison of the realistic and paranoid security noise model. The simulation parameters are selected to match the parameters used in recent LLO-CVQKD experiments [17]. Alice modulation variance $V_A$ =4 and reference pulse intensity $E_R^2/V_A$ = 100

## 3. Reference pulse attack under trusted phase noise

### 3.1 Reference pulse attack model

In CVQKD, the parameter estimation procedure allows Alice to estimate only the overall excess noise $\xi_{tot}$ value as a single parameter value. It has contributions from several individual noise sources which are practically impossible to distinguish. Consequently, the variation in the noise contribution from individual noise sources cannot be detected if the $\xi_{tot}$ is kept unchanged. We exploit this vulnerability to propose the reference pulse attack. This attack is focused on LLO-CVQKD system with trusted phase noise where the phase noise value is calibrated in advance.

The fundamental plan of the reference pulse attack is to manipulate the individual excess noise contributions without altering $\xi_{tot}$. The attack works as follows. Eve increase her amount of attack on the quantum signal, which inevitably increase the excess noise. However, at the same time, she decreases the phase estimation error noise $\xi_{esti}$, that is associated with the amplitude of the reference pulse, by an equivalent amount. Therefore, the $\xi_{tot}$ calculated in the parameter estimation process remains unchanged. As $\xi_{esti}$ is calibrated and trusted, it leaves eve to access extra information without being revealed her presence.

Without loss of generality, we consider the reference pulse attack based on a typical fibre optic implementation of a LLO-CVQKD system [17,18] as shown in the schematic diagram in Fig. 4. The reference pulse attack happens during the transmission of the pulses though the channel. At the output of Alice, Eve distinguishes and separates the signal and reference pulse into two individual channels, which can be done using a fast switch, for example. All quantum signals are transmitted through standard single mode fibre (SMF) with an attenuation factor $\alpha_{std} = 0.2 dB/km$, while the reference pulses are transmitted through low loss channel of attenuation coefficient $0 < \alpha_{low} < \alpha_{std}$. Where, $\alpha_{low} = 0$ corresponds to lossless (vacuum) channel. $\alpha_{low} = 0.14$ dB/km is currently achievable with hollow-core fibres [32]. At the input of Bob, Eve recombines the pulses into a single SMF channel.

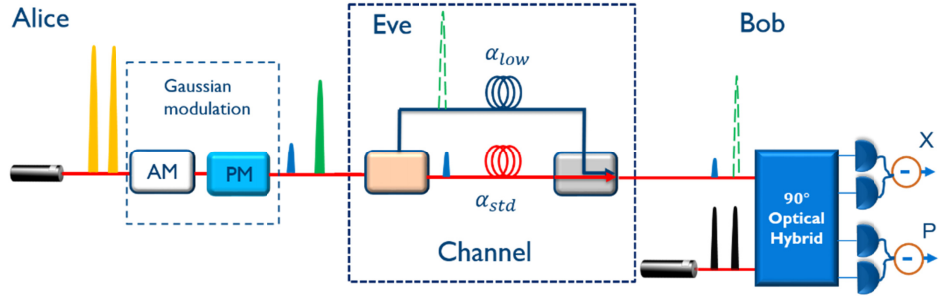

Fig. 4. Reference pulse attack schematic diagram. At the output of Alice, Eve separates the reference pulse to the low loss channel (blue) and the quantum pulse to the normal fibre (red). She recombines the pulses at the input of Bob. The dashed green lines show the less-attenuated path reference pulses.

In addition to separating the signal and reference pulses and then recombining them, Eve also performs another task, simultaneously. She increases her level of attack on the signal pulses and gains more information than usual. This will, however, increase the excess noise by $\xi_e^{attack}$ which might be detected by the users as an elevation in total excess noise $\xi_t$. As a consequence, this will lead to the actual estimation of a new Holelvo bound for Eve's information, $\chi_{BE}^{actual}$, by Alice and Bob. But, since the reference pulses propagate through low loss channel and reach Bob at an elevated amplitude, as we can see from Eq. (6), it will reduce the noise associated with the phase estimation error noise. The increase in the total excess noise is thus compensated, and Alice and Bob estimate $\chi_{BE}^{attack}$ instead of $\chi_{BE}^{actual}$ under the attack.

The additional amount of the information that Eve can steal without increasing the total excess noise is, of course, limited by the channel loss of the reference pulses. We quantify this in the following subsection. We also explored the measuring and reproducing reference pulse technique by Eve to obtain even higher reference pulse intensity. However, this process increases the uncertainty of reference pulse, hence introducing extra excess noise to system. Therefore, the effective way to achieve attack is to control the channel loss.

### 3.2 Excess noise tolerance

For the success of the attack, the increase in excess noise due to the attack on the signal must be compensated for by a simultaneous reduction in the phase noise. This requires the estimation of the noise margin available to Eve from the reduction of phase estimation error noise that is associated with the amplitude of the reference pulse. From the Eq. (6), the reduced phase estimation error noise $\xi_{esti}^{attack}$ can be evaluated as:

$$\xi_{esti}^{attack} = V_A * \frac{\chi_{tot}+1}{\frac{\eta * T_{low}}{\eta * T_{std}} E_{Ref}^2} = \xi_{esti}^{std} * \frac{T_{std}}{T_{low}} \qquad (21)$$

where $T_{std}$ is the transmittance of SMF fibre with attenuation coefficient $\alpha_{std} = 0.2$dB/km and $T_{low}$ is transmittance of the low loss reference pulse channel with an attenuation coefficient $\alpha_{low} < \alpha_{std}$. The available phase excess noise tolerance that Eve can account for by the attack can be evaluated as:

$$\xi_{tole} = \xi_{esti}^{std} - \xi_{esti}^{attack} = V_A * \frac{\chi_{tot}+1}{E_{Ref}^2} * (1 - \frac{1}{10^{(\alpha_{std}-\alpha_{low})*L/10}}) \qquad (22)$$

Here, we assume the length L of the quantum channel and reference channel are equal. Based on the value $\alpha_{low}$ of the reference channel and channel length L, Eve can increase her attack, for example, by tuning the splitting factor in intercept and resend attack, until the additional excess noise $\xi_e^{attack}$ equals $\xi_{tole}$. Hence, under the scope of reference pulse attack, the $\xi_{tot}$ estimated by Alice and Bob including calibrated phase noise is unchanged and expressed as:

$$\xi_{tot} = \xi_e + \xi_e^{attack} + \xi_{phase} - \xi_{tole} + \xi_{drift} + \xi_{AM} + \xi_{ADC} \qquad (23)$$

As explained earlier, the first two noise terms ($\xi_e + \xi_e^{attack}$) are attributed to Eve's attack on the signal while the last two terms ($\xi_{phase} - \xi_{tole}$) represent the phase noise linked to the reference pulse. Under the reference pulse attack, the quadrature values measured by Bob, Eq. (7), can be rewritten as:

$$\begin{pmatrix}\widehat{X_B^{attack}}\\\widehat{P_B^{attack}}\end{pmatrix} = \sqrt{\frac{T\eta}{2}}\left[\begin{pmatrix}\cos\widehat{\varphi_{RPA}} & \sin\widehat{\varphi_{RPA}}\\-\sin\widehat{\varphi_{RPA}} & \cos\widehat{\varphi_{RPA}}\end{pmatrix}\begin{pmatrix}X_A\\P_A\end{pmatrix} + \begin{pmatrix}X_\xi + X_\xi^{add} + X_N\\P_\xi + P_\xi^{add} + P_N\end{pmatrix}\right] + \begin{pmatrix}x_{ele}\\p_{ele}\end{pmatrix} \qquad (24)$$

Here, the phase estimator $\widehat{\varphi_{RPA}}$ is more accurate than one in Eq.(7). The most powerful impact of our attack is the applicability on every LLO-CVQKD configuration as long as phase estimation error noise is trusted. Attack creates excess noise tolerance and information leakage to Eve without being noticed. Even 100% secret key can be gained by Eve through this attack. In the following section, we estimate Eve's gain in secure key information during the attack.

### 3.3 Secret key analysis under attack

In this section, we compare the mutual information and Holevo bound among the estimated values in trusted model without attack ($I_{AB}$ and $\chi_{BE}$), with attack ($I_{AB}^{est}$ and $\chi_{BE}^{est}$) and actual information access by Eve in trusted model under attack ($\chi_{BE}^{attack}$). Based on eq. (23), the total noise estimated by Alice and Bob after attack, $\chi_t^{attack}$, is identical to the 'no' attack situation $\chi_t$ in untrusted model. This indicates that the attack, as in Eq. (13), does not affect the mutual information between Alice and Bob which can be referred to as:

$$\mathrm{I}_{AB} = \log_2 \frac{V + \chi_{tot}}{1 + \chi_{tot}} = \log_2 \frac{V + \chi_t^{attack}}{1 + \chi_t^{attack}} = \mathrm{I}_{AB}^{est} \tag{25}$$

Similarly, with pre-calibrated phase estimation error noise, Alice and Bob are unable to discover the individual change of $\xi_{esti}$ and $\xi_{drift}$ as long as the total added noise $\xi_{tot}$ remains the same. Therefore, two users still employ previous noise values for Eve's information estimation where $\chi_{het}^{est} = \chi_{het}$ and $\chi_{line}^{est} = \chi_{line}$, hence $\chi_{BE}^{est} = \chi_{BE}$. However, the actual line and detector noise terms through attack can be expressed as:

$$\chi_{het}^{attack} = \frac{[1 + (1-\eta) + 2V_{ele}]}{\eta} + T(\xi_{esti} - \xi_{tole}) \tag{26}$$

$$\chi_{line}^{attack} = \frac{1}{T} - 1 + \xi_e + \xi_{drift} + \xi_{AM} + \xi_{ADC} + \xi_e^{attack} \tag{27}$$

where excess noise $\xi_e^{attack}$ equals $\xi_{tole}$.

As a result, following the exact procedure from Eq. (14) to Eq. (20) in the trusted phase noise model, Eve's gain in information $\chi_{BE}^{attack}$ is greater than $\chi_{BE}^{est}$ due to her additional attack on the signal. This leads to $\chi_{BE}^{RPA} > \chi_{BE}^{est} = \chi_{BE}$.

Here we define the efficiency of the reference pulse attack as the ratio of the original secure information leaked to Eve with and without attack:

$$K_{eff} = \frac{\chi_{BE}^{attack} - \chi_{BE}^{est}}{I_{AB}^{est} - \chi_{BE}^{est}} \tag{28}$$

The following section reports the results of the phase noise reduction due to attack and efficiency of the attack under different LLO-CVQKD system conditions.

## 4. Performance Analysis

### 4.1 Excess noise tolerance

We investigate the attack performance as a function of various reference channel attenuation factors, $\alpha_{low}$, ranging from the vacuum channel (0dB/km) to conventional SMF (0.2dB/km). We also consider a practical scenario where the reference channel uses recently reported hollow-core fibre of 0.14dB/km attenuation coefficient [30]. The simulation parameters are designated as the same as in Sec 2.2 to provide optimal performance. Values are expressed in shot noise units.

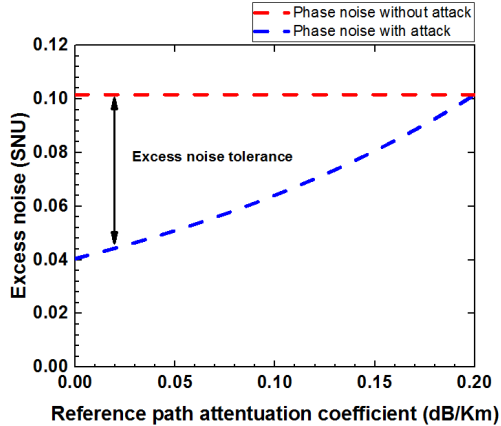

Fig. 5. Phase noise with (blue dashed line) and without (red dashed line) attack variation at different reference path attenuation coefficient. Excess noise tolerance is generated by attack. Simulations are performed in the collective attack and $V_A=4$, $E_{Ref}^2/V_A = 100$ and $L = 20$ km.

According to the Eq. (22), the phase excess noise tolerance increases with increase in transmission distance, decrease in reference path's attenuation coefficient, and drop with the reference pulse intensity. However, the too large reference pulse intensity also has negative impact on key rate as we discussed previously. In normal case, Alice and Bob will pick the optimal reference pulse intensity without the consideration of our attack. Consequently, we investigate our attack performance at the optimal reference intensity. Fig. 5 shows the variation in phase excess noise tolerance with different attenuation coefficients of the reference channel for a 20 km LLO-CVQKD link at $\frac{E_R^2}{V_A} = 100$. Attack works for any channel attenuation factor and the phase noise reduces by 60.2% and 24.1% from the initial value, for $\alpha_{low} = 0$ dB/km and 0.14dB/km reference channels, respectively.

Fig.6 shows the variation in mutual information as a function of transmission distance, with optimal reference pulse intensity, under trusted phase noise model. The $\chi_{BE}^{attack}$ values are only evaluated for theoretical minimum $\alpha_{low} = 0$ dB/km (blue dashed line) and practical minimum 0.14 dB/km (red dashed line). $\chi_{BE}^{est}$, is evaluated with reference path attenuation coefficient $\alpha_{std} = 0.2$ dB/km (purple solid line) under the notion that the attack is undetectable to Alice and Bob so channel attenuation coefficient for signal and reference channels are equal. The mutual information between Alice and Bob, $I_{AB}$ (black solid line), monotonically decreases as a function of channel length. The secure region under attack is labelled as the truly secure region while the yellow region is the insecure region estimated by Alice and Bob with and without attack. The white region referred as the attack induced insecure region where Eve is successful in her attack without revealing her presence. At relatively long distances, this grows rapidly which indicates an increase in attack efficiency. We need to emphasize that the $K_{eff}$ value varies with system parameters and a 100% insecure key situation can be achieved, especially at long transmission distance. It is noted that the achievable secure distance under trusted noise model without attack is 99.1km. Attack with lossless reference channel, unity attack efficiency reduces it to 23.8km. In more practical sense, the maximum transmission distance is reduced to 32.0km in reference channel with 0.14dB/km loss. This indicates significant attack impact in the practical LLO CVQKD systems.

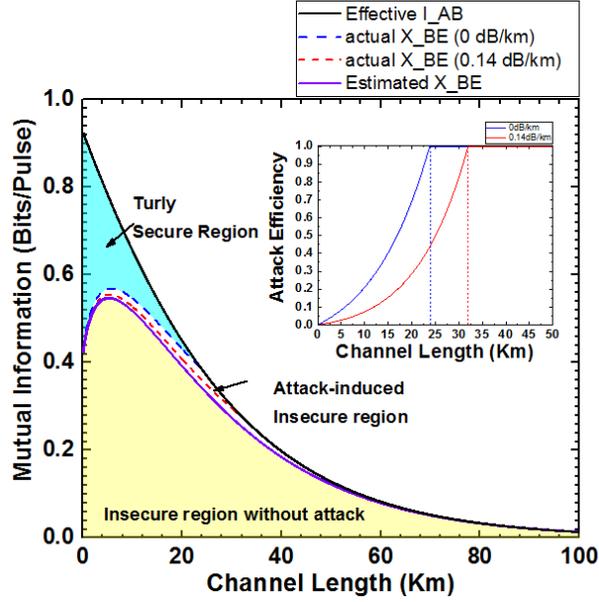

Fig. 6. Mutual information in realistic model vs channel lengths. Insecure region expands with increase in transmission distance. Zero-loss (vacuum) channel maximize the information gain by Eve. 100% attack efficiency is achieved from 23.8km and 32.0 km for 0 and 0.14 dB/km attenuation factors.

Fig.7 compares the performance of the untrusted phase noise LLO-CVQKD system performance with the trusted LLO-CVQKD system under reference pulse attack, at 20km transmission distance. At 20km there is no secure key generation possible for untrusted noise LLO-CVQKD system. However, we use this situation to show merits of trusted noise model. In Fig.7(a), for untrusted phase noise model, it shows the $\chi_{BE}$ is greater then $I_{AB}$ that lead to zero key rate. On the other hand, for trusted phase noise model, Fig.7(b), since the calibrated $\xi_{esti}$ is excluded from excess noise estimation, the mutual information between Alice and Bob $I_{AB}$ is always stays higher than Eves' information upper bound. However, the attack on reference pulse increase the Holevo bound for Eve's information which estimated by Alice and Bob as $\chi_{BE}^{est}$ (blue dashed line) but, accounts to $\chi_{BE}^{attack}$ (red dashed line). This creates an attack-induces insecure region. Quantitatively, at a 20 km distance and for 0dB/km attenuation coefficient for the reference channel, Eve can extract 48.6% of the secret key without the attack being identified. For a feasible 0.14 dB/km attenuation coefficient, Eve can achieve 23.2% additional information. The attack reduces the key rate, however, there exist at secure region that sill can be exploited for key generation. This indicates the potentiality of trusting the phase noise to improve key rate as well as transmission distance. The feasibility of trusted phase noise model is investigated in next section.

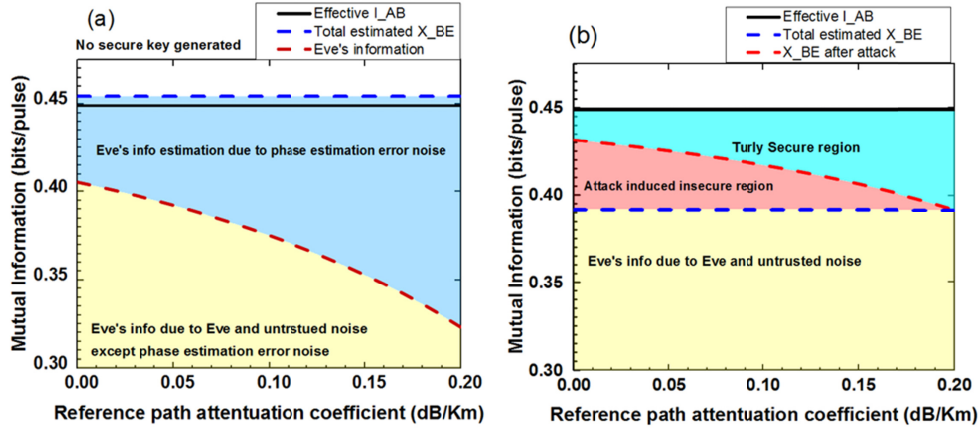

Fig.7. Mutual information variations for different reference path attenuation coefficients. Simulations are performed for collective attack, with (a) untrusted phase noise assumption (b) trusted phase noise assumption.

*4.2 Feasibility of the trusted phase noise model*

The trusted phase noise assumption contributes to an extended transmission distance and improved secret key rate, even under reference pulse attack. However, in order to exploit the positive key rate under the attack, we need to propose a countermeasure in the trusted noise model. This involves the estimation of Eve's tolerable noise $\xi_{tole}$ to attack the signal while using lossless channel for the reference pulse transmission. By accounting $\xi_{tole}$ to the channel loss, one can upper bound Eve's information for each signal transmission distance. For example, for 20km signal transmission distance, this corresponds to the $\chi_{BE}$ at 0 dB/km reference path attenuation.

The system performance varies with the calculated $\xi_{tole}$ value. $\xi_{tole}$ is dependent on the reference path attenuation coefficient, distance and reference pulse intensity. As a consequence, we choose $\alpha_{low}$ as 0.14 dB/km and 0dB/km at different optimal reference pulse intensity to show the validation of the countermeasure and the trusted phase noise system performance. The modulation variance and other parameters are chosen as the previous optimal values. Adapting this strategy, Fig.8 shows the enhancement in key rate as well as transmission distance for various reference pulse intensity. More importantly, after updating the noise model, the resultant secure transmission distance and secret key rate are still much better than conventional untrusted phase noise model, especially at relatively weak reference pulse situations. Specifically, under the optimal $E_{ref}^2/V_A = 100$, trusted phase noise model can achieve 43.5% longer transmission distance and roughly 9.5 times elevation in the key rate at the current technological state of art - 0.14dB/km fibre. Also, at the most pessimistic case, 0dB/km fibre, the countermeasure is still valid to achieve improved performance than untrusted phase noise model. It worth to mention that the further performance improvement can be accomplished by employing higher extinction ratio amplitude modulators and high bit count analogue to digital conversion devices.

Another way to protect against the attack is to monitor the instantaneous amplitude of the reference pulse and calibrate phase noise in real time, so that Eve's information can be accordingly upgraded. If the $\chi_{BE}$ is greater than $I_{AB}$, obtained from parameter estimation of the signal, Alice and Bob can simply stop communication. This will lead to further performance improvement with higher software and hardware requirement.

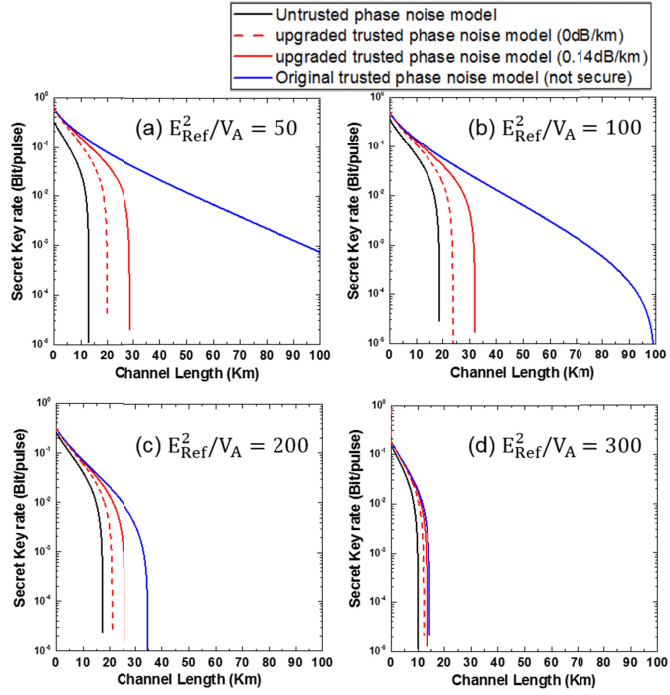

Fig.8. Updated trusted phase noise model performance comparison at different reference pulse intensities. The black line represents the key rate performance under untrusted phase noise model. Blue line shows the trusted phase noise model performance without considering loophole we discovered in this paper. The red solid line and dashed line respectively demonstrate the resultant secure key performance at practically the lowest fibre loss 0.14dB/km and the idealistic lossless fibre 0dB/km after upgrading noise model in this paper. The different pulse intensity ratios are applied in (a) 50 (b) 100 (c) 200 and (d) 300. We choose all other parameters are identical to the values shown in Sec II.C. The upgraded noise model provides enhanced secure key under trusted phase noise model to untrusted case in any reference pulse intensity.

## 5. Conclusion

In this paper, we have, revealed a security threat in the realistic LLO-CVQKD system with trusted phase noise called the reference pulse attack. This comprises two simultaneous tasks: (a) where Eve selectively switches reference pulses to a low loss channel; and (b) she attacks the quantum signals. We have shown that the phase estimation error noise can be tuned by adjusting the reference channel attenuation coefficient which provides freedom to Eve to acquire extra information from the quantum signals. However, even under reference pulse attack, the trusted noise model of LLO-CVQKD system shows higher security key rate and longer distance than untrusted LLO-CVQKD system. More importantly, this estimation error exists in all reference pulse protocols. Therefore, the attack could affect all LLO CVQKD. In the future, we are going to investigate other parameters relating to the realistic security scenario for a more vigorous security analysis and possible detection of side channels.

**Funding.** EPSRC Quantum Communication Hub EP/M013472/1
**Acknowledgement.** We would like to thank Hoi-Kwong Lo, Romain Allèaume and Bing Qi for their important comments and useful suggestions.

**References**

1. C. H. Bennett and G. Brassard, "Quantum Cryptography: Public Key Distribution and Coin Tossing," in Proceedings of IEEE International Conference on Computers, Systems and Signal Processing (IEEE, New York, 1984), pp. 175–179.
2. A. K. Ekert, Quantum Cryptography Based on Bell's Theorem, Phys. Rev. Lett. 67, 661 (1991).
3. V. Scarani, H. Bechmann-Pasquinucci, N. J. Cerf, M. Dušek, N. Lütkenhaus, and M. Peev, The Security of Practical Quantum Key Distribution, Rev. Mod. Phys. 81, 1301 (2009).
4. H.-K. Lo, M. Curty, and K. Tamaki, Secure Quantum Key Distribution, Nat. Photonics 8, 595 (2014).
5. Grosshans, Frédéric, et al. Quantum key distribution using gaussian-modulated coherent states. Nature 421.6920 (2003).
6. Leverrier, Anthony, and Philippe Grangier. Unconditional security proof of long-distance continuous-variable quantum key distribution with discrete modulation. Phy. Rev. Lett. 102.18 (2009).
7. Weedbrook, Christian, et al. Gaussian quantum information. Rev. Mod. Phys. 84.2 (2012): 621.
8. Diamanti, Eleni, and Anthony Leverrier. Distributing secret keys with quantum continuous variables: principle, security and implementations. Entropy 17.9 (2015).
9. Huang, Duan, et al. High-speed continuous-variable quantum key distribution without sending a local oscillator. Opt. Lett. 40.16 (2015).
10. Chi, Yue-Meng, et al. A balanced homodyne detector for high-rate Gaussian-modulated coherent-state quantum key distribution. New J. Phys.13.1 (2011).
11. Huang, Duan, et al. A wideband balanced homodyne detector for high speed continuous variable quantum key distribution systems. Qcrpyt, Waterloo, Canada. 2013.
12. Navascués, Miguel, Frédéric Grosshans, and Antonio Acin. Optimality of Gaussian attacks in continuous-variable quantum cryptography. Phys. Rev. Lett. 97.190502 (2006).
13. Leverrier, Anthony, Frédéric Grosshans, and Philippe Grangier. Finite-size analysis of a continuous-variable quantum key distribution. Phys. Rev. A 81.6 062343 (2010).
14. Ma, Xiang-Chun, et al. Enhancement of the security of a practical continuous-variable quantum-key-distribution system by manipulating the intensity of the local oscillator. Phys. Rev. A 89.3 032310 (2014).
15. Huang, Jing-Zheng, et al. Quantum hacking of a continuous-variable quantum-key-distribution system using a wavelength attack. Phys. Rev. A 87.6 062329 (2013).
16. Qin, Hao, Rupesh Kumar, and Romain Alléaume. Quantum hacking: Saturation attack on practical continuous-variable quantum key distribution. Phys. Rev. A 94.1 012325 (2016).
17. Qi, Bing, et al. Generating the local oscillator "locally" in continuous-variable quantum key distribution based on coherent detection. Phys. Rev. X 5.4 041009 (2015).
18. Soh, Daniel BS, et al. Self-referenced continuous-variable quantum key distribution protocol. Phys. Rev. X 5.4 041010 (2015)
19. Huang, Duan, et al. High-speed continuous-variable quantum key distribution without sending a local oscillator. Opt. Lett. 40.16 (2015).
20. Kleis, S., Rueckmann, M., & Schaeffer, C. G. Continuous variable quantum key distribution with a real local oscillator using simultaneous pilot signals. Opt. Lett.,42(8) (2017).
21. Marie, A., & Alléaume, R.. Self-coherent phase reference sharing for continuous-variable quantum key distribution. Phys. Rev. A, 95(1), 012316 (2017).
22. Kleis, Sebastian, Max Rueckmann, and Christian G. Schaeffer. "Continuous variable quantum key distribution with a real local oscillator using simultaneous pilot signals." Optics letters 42.8, 1588-1591, (2017).
23. Wang, Tao, et al. "Pilot-multiplexed continuous-variable quantum key distribution with a real local oscillator." Physical Review A 97.1 012310 (2018).
24. Laudenbach, Fabian, et al. "Continuous‐Variable Quantum Key Distribution with Gaussian Modulation—the Theory of Practical Implementations." Advanced Quantum Technologies 1.1 1800011 (2018).
25. Qi, Bing, et al Noise Analysis of simultaneous quantum key distribution and classical communication scheme using a true local oscillator, Phys.Rev.Applied 9, 054008
26. Lodewyck, Jerome, et al. Quantum key distribution over 25km with an all-fiber continuous variable system. Phys.Rev.A 75, 042305
27. Holevo, Alexander Semenovich. Bounds for the quantity of information transmitted by a quantum communication channel. Problemy Peredachi Informatsii 9.3 (1973)
28. Shannon, C.E. A mathematical theory of communication. Bell Syst. Tech. J. (1948)
29. Von Neumann, John.Mathematical foundations of quantum mechanics. No. 2. Princeton university press, 1955.
30. Adesso, Gerardo, Alessio Serafini, and Fabrizio Illuminati. Extremal entanglement and mixedness in continuous variable systems. Phys. Rev. A 70.2 022318 (2004).
31. Fossier, Simon, et al. Improvement of continuous-variable quantum key distribution systems by using optical preamplifiers. J. Phys. B 42.11 114014 (2009).
32. Tamura, Yoshiaki, et al. "Lowest-ever 0.1419-dB/km loss optical fiber. OFC Conference. Optical Society of America, 2017.